\documentclass[prc,twocolumn,showpacs,floatfix,nofootinbib,superscriptaddress]{revtex4}

\usepackage{amsmath,amssymb,graphicx,bm}

\newcommand{\beqn}{\begin{equation}}
\newcommand{\eeqn}{\end{equation}}
\newcommand{\bea}{\begin{eqnarray}}
\newcommand{\eea}{\end{eqnarray}}

\newcommand{\la}{\langle}
\newcommand{\ra}{\rangle}

\newcommand{\nmax}{$N_{\rm max}$}
\newcommand{\hw}{\hbar\Omega}

\newcommand{\adag}{a^\dagger}

\begin{document}


\title{Evolution of Nuclear Many-Body Forces \\
with the Similarity Renormalization Group}

\author{E.D.\ Jurgenson}
\affiliation{Department of Physics, The Ohio State University, 
Columbus, OH\ 43210, USA}

\author{P.\ Navr\'atil}
\affiliation{Lawrence Livermore National Laboratory, P.O. Box
808, L-414, Livermore, CA\ 94551, USA}

\author{R.J.\ Furnstahl}
\affiliation{Department of Physics, The Ohio State University, 
Columbus, OH\ 43210, USA}

%
%


\begin{abstract}
The first practical method to evolve many-body nuclear forces to 
softened form using the Similarity Renormalization Group (SRG) in
a harmonic oscillator basis is demonstrated.  When applied to
$^4$He calculations, the two- and three-body oscillator matrix
elements yield rapid convergence of the ground-state energy with
a small net contribution of the induced four-body force. 
\end{abstract}
\smallskip

\maketitle


A major goal of nuclear structure theory is to make quantitative
calculations of low-energy nuclear observables starting from microscopic
internucleon forces. Chiral effective field theory ($\chi$EFT)
provides a systematic construction of these forces, including a
hierarchy of many-body forces of decreasing
strength~\cite{Epelbaum:2008ga}. Renormalization group (RG)
methods can be used to soften the short-range repulsion and
short-range tensor components of the initial chiral interactions
so that convergence of nuclear structure calculations is
greatly accelerated~\cite{Bogner:2006vp,Bogner:2007rx}. The
difficulty is that these transformations (or any other softening
transformations) change the short-range many-body forces.   
To account for these changes, we present in this
letter the first consistent evolution of three-body
forces by using the Similarity Renormalization Group 
(SRG)~\cite{Glazek:1993rc,Wegner:1994,Kehrein:2006,Bogner:2006srg,Bogner:2007srg},
which offers a technically simpler approach to evolving
many-body forces than other RG formulations. 
Our results show that both the many-body hierarchy of $\chi$EFT
and the improved convergence properties are preserved.

The SRG is a series of unitary
transformations of the free-space Hamiltonian 
$(H \equiv H_{\lambda=\infty}$), 
\beqn
H_{\lambda} = U_{\lambda} H_{\lambda=\infty}U_{\lambda}^{\dagger} \;,
\label{eq:Hs}
\eeqn
labeled by a momentum parameter $\lambda$ that runs 
from $\infty$ toward zero,
which keeps track of the
sequence of Hamiltonians ($s = 1/\lambda^4$ 
has been used elsewhere~\cite{Bogner:2006srg,Bogner:2007srg}).
These transformations are implemented as a flow equation in $\lambda$
(in units where $\hbar^2/M = 1$),
\beqn
 \frac{dH_{\lambda}}{d\lambda} = -\frac{4}{\lambda^5}
                       [[T,H_{\lambda}],H_{\lambda}]  \;,
 \label{eq:flow}
\eeqn
whose form guarantees that the $H_\lambda$'s are unitarily 
equivalent~\cite{Kehrein:2006,Bogner:2006srg}.

The appearance of the nucleon kinetic energy $T$ in
Eq.~\eqref{eq:flow} leads to high- and low-momentum parts of 
$H_{\lambda}$ being decoupled, which means softer and more
convergent potentials~\cite{Jurgenson:2007td}.
This is evident in a partial-wave momentum basis, 
where matrix elements $\la k | H_\lambda | k' \ra$ 
connecting states with (kinetic) energies differing by more than
$\lambda^2$ are suppressed by $e^{-(k^2-k'{}^2)^2/\lambda^4}$ factors 
and therefore the states decouple
as $\lambda$ decreases. 
(Decoupling also results from replacing $T$ in
Eq.~\eqref{eq:flow} with other 
generators~\cite{Kehrein:2006,Bogner:2006srg,Glazek:2008pg,Anderson:2008mu}.)
The
optimal range for $\lambda$ is not yet established and also
depends on the system, but experience with SRG and other
low-momentum potentials suggest  that running to about $\lambda =
2.0\,\mbox{fm}^{-1}$  
is a good compromise between improved
convergence from decoupling and the growth of induced
many-body interactions~\cite{Jurgenson:2007td}.
(Also, differences between using $T$ and the diagonal of $H_\lambda$
in Eq.~\eqref{eq:flow}, which can be very important in some
situations~\cite{Glazek:2008pg}, 
are negligible in this $\lambda$ range.)

To see how the two-, three-, and higher-body potentials
are identified, it is useful to decompose $H_\lambda$ in second-quantized
form.  Schematically (suppressing indices and sums),
\beqn
  H_\lambda = \la T \ra \adag a  + \la V_\lambda^{(2)} \ra \adag\adag a a
       + \la V_\lambda^{(3)} \ra \adag\adag\adag a a a + \cdots
       \;,
       \label{eq:2ndquant}
\eeqn
where $\adag$, $a$ are creation and destruction operators
with respect to the vacuum in some (coupled) single-particle basis. 
This \emph{defines} $\la T \ra$, $\la V_\lambda^{(2)} \ra$,
$\la V_\lambda^{(3)} \ra$, \ldots as the one-body,
two-body, three-body, \ldots
matrix elements at each $\lambda$. 
Upon evaluating
the commutators in Eq.~\eqref{eq:flow} using $H_\lambda$ from
Eq.~\eqref{eq:2ndquant}, we see that even if initially there are only
two-body potentials, higher-body potentials are generated with
each step in $\lambda$. 
Thus, when applied in an $A$-body subspace, the SRG will ``induce''
$A$-body forces. 
But we also see that $\la T \ra$ is
fixed, $\la V_\lambda^{(2)} \ra$ is determined only in the $A=2$
subspace with no dependence on $\la V_\lambda^{(3)} \ra$, $\la V_\lambda^{(3)} \ra$ is determined  in $A=3$ given $\la
V_\lambda^{(2)} \ra$, and so on.

Since only the Hamiltonian enters
the SRG evolution equations, there are no difficulties from having
to solve T matrices in all channels for different $A$-body
systems.
However, in a momentum basis the presence of spectator nucleons
requires solving separate equations for each set
of $\la V^{(n)}_\lambda \ra$ matrix elements.
In Refs.~\cite{Bogner:2007qb,Jurgenson:2008jp}, a diagrammatic
approach is introduced to handle this decomposition.
But while it is natural to solve Eq.~\eqref{eq:flow} in momentum
representation, it is an operator equation so we can use any
convenient basis. 
Here we evolve in a \emph{discrete} basis, where spectators
are handled without a decomposition and induced many-body forces
can be directly identified.  
Having chosen such a basis, we obtain coupled
first-order differential equations for the matrix elements of the
flowing Hamiltonian $H_\lambda$, where the right side
of Eq.~\eqref{eq:flow} is evaluated
using simple matrix multiplications.

Our calculations  are performed in the Jacobi coordinate
harmonic oscillator (HO) basis of the No-Core Shell Model
(NCSM)~\cite{Navratil:2009ut}. This is a translationally
invariant, anti-symmetric basis for each $A$, with a complete set
of states up to a maximum excitation of \nmax$\hbar\Omega$ above
the minimum energy configuration, where $\Omega$ is the harmonic
oscillator parameter. The procedures used here build directly on
Ref.~\cite{Jurgenson:2008jp}, which presents a one-dimensional
implementation of our approach along with a general analysis
of the evolving many-body hierarchy.

\begin{table*}[tbh-]
\caption{\label{tab:one}Definitions of the various calculations.}
\begin{ruledtabular}
\begin{tabular}{crlc}
 & NN-only & No initial NNN interaction and do not keep
              NNN-induced interaction. &  \\
 & NN + NNN-induced & No initial NNN interaction but keep the
         SRG-induced NNN interaction. &  \\
 & NN + NNN & Include an initial NNN interaction \emph{and} keep
         the SRG-induced NNN interaction. &  \\
\end{tabular}
\end{ruledtabular}
\end{table*}

We start by evolving $H_\lambda$ in the $A=2$ subsystem, which completely
fixes the two-body matrix elements $\la V_\lambda^{(2)}\ra$.
Next, by evolving $H_\lambda$ in the $A=3$ subsystem we determine the
combined two-plus-three-body matrix elements.
We can isolate the three-body matrix elements
by subtracting the evolved  $\la V_\lambda^{(2)}\ra$ elements
in the $A=3$ basis~\cite{Jurgenson:2008jp}. Having obtained
the separate NN and NNN matrix elements, 
\emph{we can apply them unchanged to any nucleus}. 
We are also free to include any initial
three-nucleon force in the initial Hamiltonian without
changing the procedure. 
If applied to $A \geq 4$, four-body (and higher) forces will not
be included and
so the transformations will be only approximately unitary. 
The questions to be addressed
are whether the decreasing hierarchy of many-body forces is
maintained and whether the induced four-body contribution
is unnaturally large.  
We summarize in
Table~\ref{tab:one} the different calculations to be made for 
$^3$H and $^4$He to confront these questions.

The initial ($\lambda = \infty$) NN potential used here
is the 500\,MeV N$^3$LO interaction from Ref.~\cite{N3LO}.  The
initial NNN potential is the N$^2$LO
interaction~\cite{Epelbaum:2002vt} 
in the local form of Ref.~\cite{Navratil:07} 
with constants fit to the average of triton and $^3$He binding 
energies and to triton beta decay  according to
Ref.~\cite{Gazit:2008ma}.
We expect similar results from
other initial interactions because the SRG drives them toward
near universal form; a survey will be given in
Ref.~\cite{Jurgenson:2009}. NCSM calculations with these initial
interactions and the parameter set in Table~I of
Ref.~\cite{Gazit:2008ma} yield energies of
$-8.473(4)\,$MeV for $^3$H and $-28.50(2)\,$MeV for $^4$He
compared  with $-8.482\,$MeV and $-28.296\,$MeV from experiment,
respectively. So there is a 20\,keV uncertainty in the
calculation of $^4$He from incomplete convergence and a
200\,keV discrepancy with experiment. 
The latter is consistent with the omission of three- and four-body
chiral interactions at N$^3$LO.
These provide a scale for
assessing whether induced four-body contributions are important
compared to other uncertainties.


\begin{figure}[thb]
\includegraphics*[width=7.7cm]{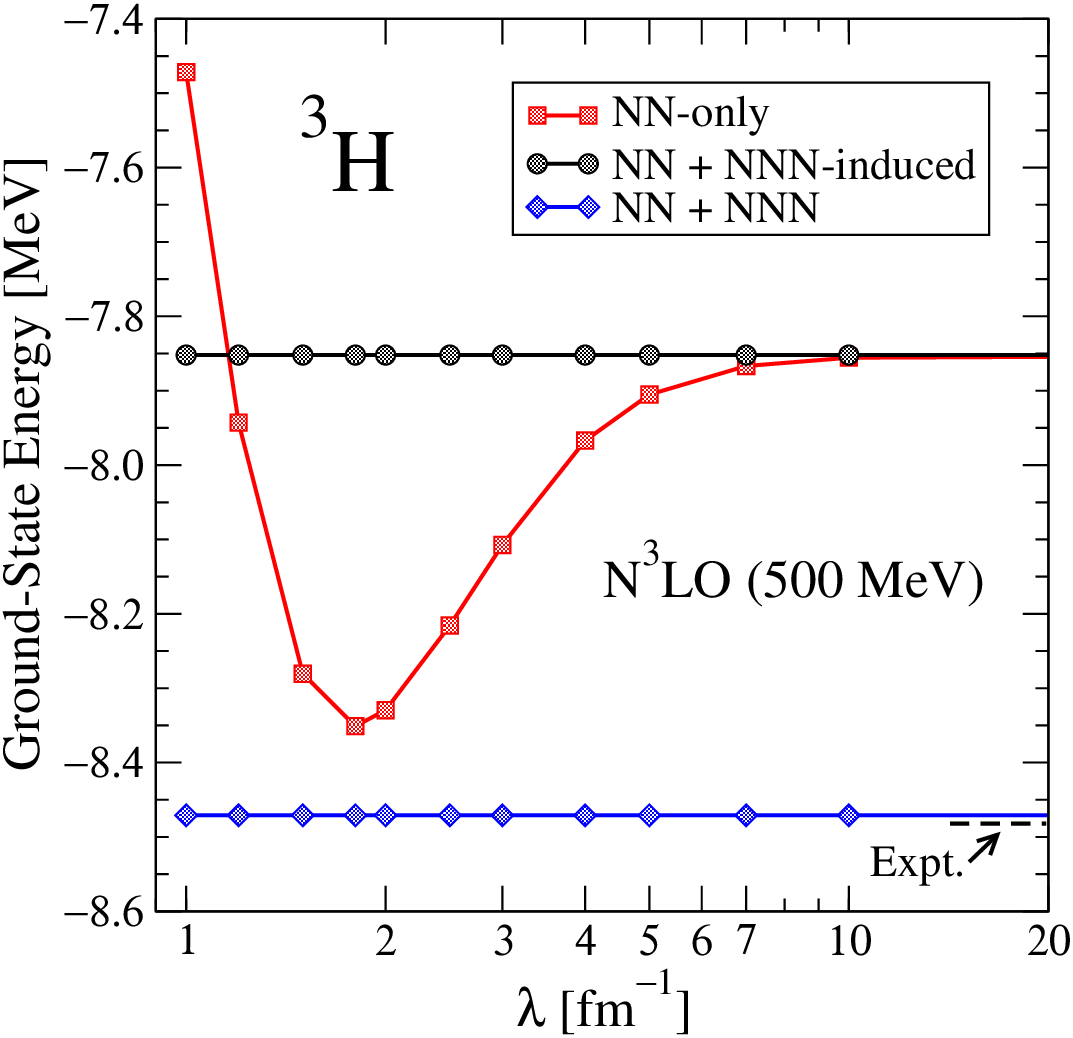}
\caption{(Color online) Ground-state energy of $^3$H as a
function of the SRG evolution parameter, $\lambda$. See
Table~\ref{tab:one} for the nomenclature of the curves.}
\label{fig:h3_srg}
\end{figure}

\begin{figure}[htb]
\includegraphics*[width=7.7cm]{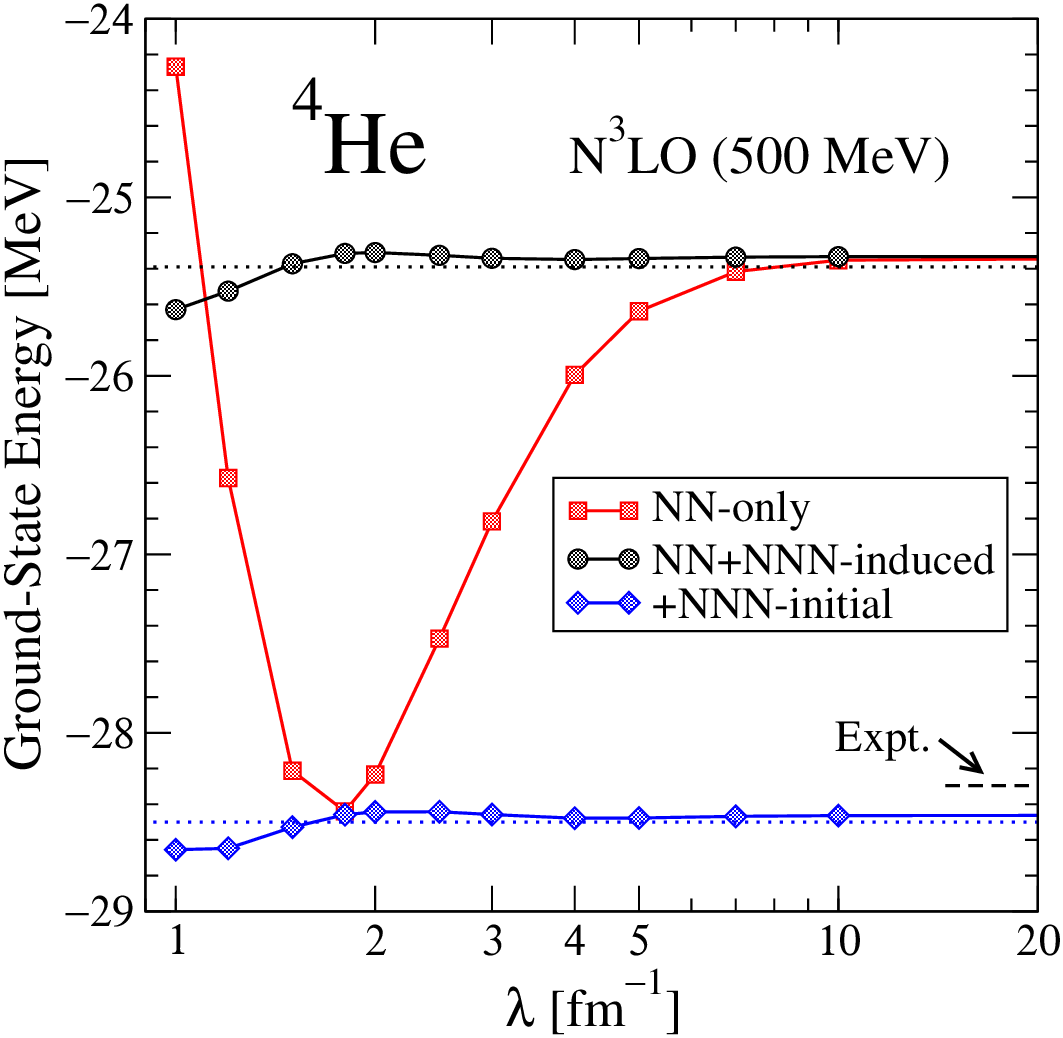}
\caption{(Color online) Ground-state energy of $^4$He as a
function of the SRG evolution parameter, $\lambda$. See
Table~\ref{tab:one} for the nomenclature of the curves.}
\label{fig:he4_srg}
\end{figure}

In Fig.~\ref{fig:h3_srg}, the ground-state energy of the triton 
is plotted as a function of the flow parameter $\lambda$. 
Evolution is from $\lambda = \infty$, which is the
initial (or ``bare'') interaction, toward $\lambda = 0$. We use
$N_{\rm max} = 36$ and $\hbar\Omega = 28\,$MeV, for which all
energies are converged to better than 10\,keV. We first consider
an NN interaction with no initial NNN (``NN-only''). 
If $H_\lambda$ is evolved only in an
$A=2$ system, higher-body induced pieces are lost.  The resulting
energy calculations will only be approximately unitary for $A>2$
and the ground-state energy  will vary with $\lambda$
(squares). Keeping the induced NNN yields a flat line
(circles), which implies an exactly unitary transformation; the
line is equally flat if an initial NNN is included (diamonds).
Note that the net induced three-body is comparable to the initial
NNN contribution and thus is of natural size.

In Fig.~\ref{fig:he4_srg}, we examine the SRG evolution in
$\lambda$ for $^4$He with $\hbar\Omega = 36\,$MeV. 
The $\la V_\lambda^{(2)}\ra$ and $\la V_\lambda^{(3)}\ra$ matrix elements
were evolved in $A=2$ and $A=3$ with $N_{\rm max} = 28$ and 
then truncated to $N_{\rm max} = 18$ at each $\lambda$
to diagonalize $^4$He.
The NN-only curve has a similar shape as
for the triton. In fact, this pattern of variation has been
observed in all SRG calculations of light nuclei~\cite{Bogner:2007rx}. 
When the induced NNN is included, the
evolution is close to unitary  and the pattern only depends slightly on an
initial NNN interaction. In both cases the dotted line
represents the converged value for the initial Hamiltonian. At
large $\lambda$, the discrepancy is due to a lack of convergence at
$N_{\rm max} = 18$, but at $\lambda < 3\,\mbox{fm}^{-1}$ SRG
decoupling takes over and the discrepancy is due to short-range induced
four-body forces, which therefore contribute about 50\,keV net
at $\lambda = 2\,\mbox{fm}^{-1}$. 
This is small compared to the rough estimate in Ref.~\cite{Rozpedzik:2006yi}
that the contribution
from the long-ranged part of the N$^3$LO four-nucleon force 
to $^4$He binding is of order a few hundred keV.  
If needed, we could evolve 4-body matrix elements in $A=4$ and
will do so when nuclear structure codes can accomodate them.


\begin{figure}[htb]
\includegraphics*[width=7.7cm]{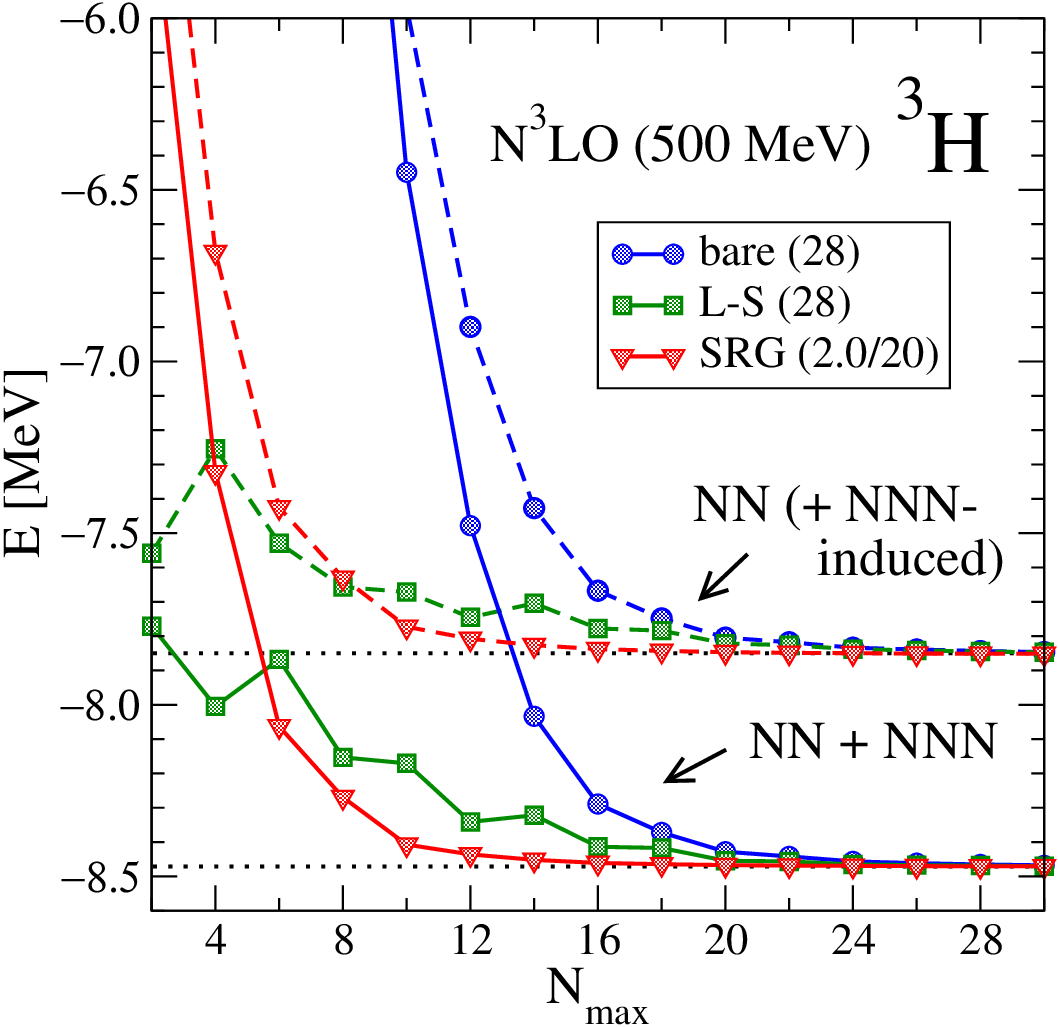}
\caption{(Color online) Ground-state energy of $^3$H as a
  function of the basis size \nmax\ for an N$^3$LO NN 
  interaction~\cite{N3LO} with and without an initial NNN
  interaction~\cite{Epelbaum:2008ga,Gazit:2008ma}.  Unevolved
  (``bare'') and Lee-Suzuki (L-S) results with $\hbar\Omega =
  28\,$MeV are compared with SRG at $\hbar\Omega = 20\,$MeV
  evolved to $\lambda = 2.0\,\mbox{fm}^{-1}$. }
\label{fig:h3_convergence}
\end{figure}

\begin{figure}[thb]
\includegraphics*[width=7.7cm]{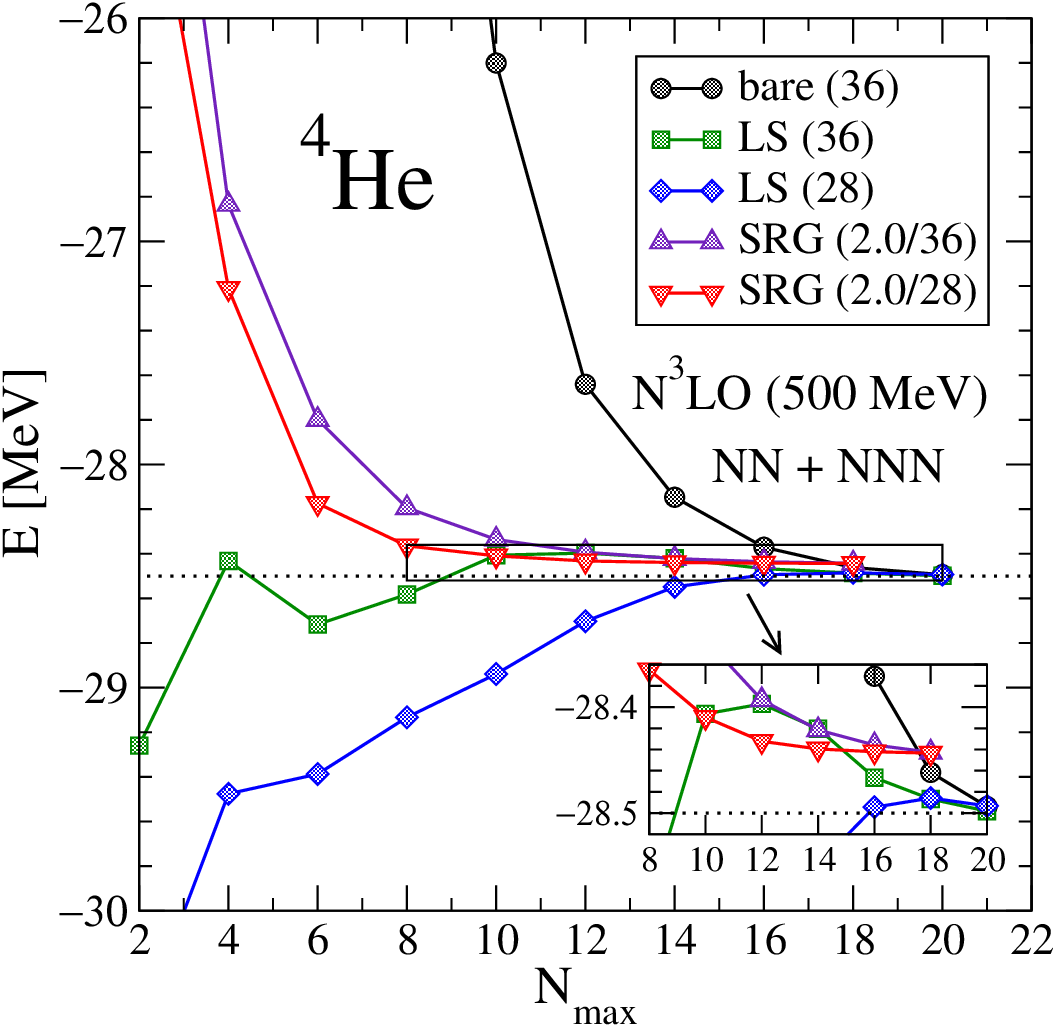}
\caption{(Color online) Ground-state energy of $^4$He as a
  function of the basis size \nmax\ for an N$^3$LO NN 
  interaction~\cite{N3LO} with an initial NNN
  interaction~\cite{Epelbaum:2008ga,Gazit:2008ma}.  Unevolved
  (bare) results are compared with Lee-Suzuki (L-S) 
  and SRG evolved to $\lambda = 2.0\,\mbox{fm}^{-1}$ 
  at $\hbar\Omega = 28$ and $36\,$MeV.}
\label{fig:he4_convergence}
\end{figure}

In Fig.~\ref{fig:h3_convergence}, we show the triton ground-state
energy as a function of the oscillator basis size, \nmax, for
various calculations.  The lower (upper) curves are with
(without) an initial three-body force (see Table~\ref{tab:one}).
The convergence of the bare interaction is compared  with the SRG
evolved to $\lambda = 2.0\,\mbox{fm}^{-1}$. The oscillator
parameter $\hw$ in each case was chosen roughly to optimize the
convergence of each Hamiltonian. (As $\lambda$ decreases, so does
the optimal $\hw$.) We also compare to a Lee-Suzuki (L-S)
effective interaction, which has been used in the NCSM to greatly
improve convergence~\cite{Nogga:2005hp,Navratil:2007we}. These
effective interactions result from unitary transformations within
the model space of a given nucleus, in contrast to the free-space
transformation of the SRG, which yields nucleus-independent
matrix elements.

The SRG calculations are variational and converge smoothly and
rapidly from above with or without an initial three-body force.
The dramatic improvement in convergence rate compared to the initial
interaction is seen even though
the  $\chi$EFT interaction is relatively soft. Thus, once
evolved, a much smaller \nmax\ basis is adequate for a desired
accuracy and extrapolating in \nmax\ is also feasible.

Figure~\ref{fig:he4_convergence} illustrates for $^4$He the same
rapid convergence with \nmax\ of an SRG-evolved interaction. 
However, in this case the
asymptotic value of the energy differs slightly because of the
omitted induced four-body contribution. (The SRG-evolved
asymptotic values for different $\hbar\Omega$ 
differ by only 10\,keV, so the gap 
between the converged bare/L-S results 
and the SRG results is dominated by the induced NNNN
rather than incomplete convergence).
Convergence is even faster for lower $\lambda$
values~\cite{Jurgenson:2009}, ensuring a
useful range for the analysis of few-body systems.
However, because of the strong density dependence of four-nucleon forces,
it will be important to monitor the size of the induced four-body 
contributions for heavier nuclei and nuclear matter.

\begin{figure}[t]
\includegraphics*[width=7.7cm]{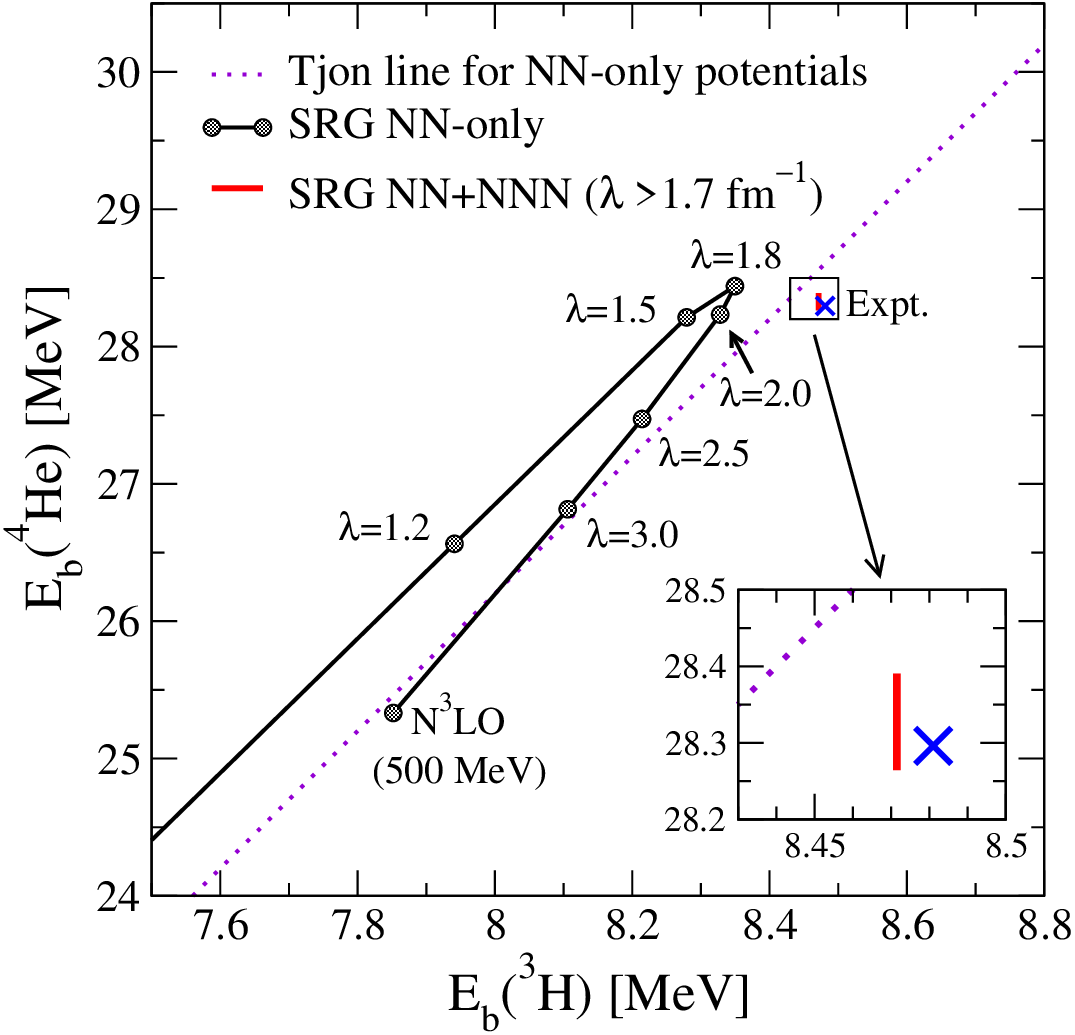}
\caption{(Color online) Binding energy of the alpha particle vs.\
  the binding energy of the triton. 
  The Tjon line from phenomenological NN potentials
  (dotted) is compared with the trajectory of SRG energies
  when only the NN interaction is kept (circles).  When the
  initial and induced NNN interactions are included, the
  trajectory lies close to experiment for $\lambda >
  1.7\,\mbox{fm}^{-1}$ (see inset). }
\label{fig:tjon_line}
\end{figure}

The impact of evolving the full three-body
force is neatly illustrated
in Fig.~\ref{fig:tjon_line}, where the binding energy of $^4$He is
plotted against the binding energy of $^3$H.  The experimental
values of these quantities, which are known to a small fraction
of a keV,  define only a point in this plane (at the center of
the X, see inset).    
The SRG NN-only results trace out a trajectory in the plane that
is analogous to the well-known Tjon line (dotted), which is the
approximate locus of points for  phenomenological potentials fit 
to NN data but not including NNN~\cite{Vlowk3N}.
In contrast,
the short trajectory of
the SRG with the NN + NNN interaction (shown for $\lambda \geq
1.8\,\mbox{fm}^{-1}$) highlights the small variations from the 
omitted four-nucleon force. Note that a
trajectory plotted for NN+NNN-induced calculations 
would be a similarly small line at the N$^3$LO NN-only point.


In summary, we have demonstrated a practical method to use the
SRG to evolve NNN (and higher many-body) forces in a
harmonic oscillator basis.
Calculations of $A \leq 4$ nuclei including NNN show the same
favorable convergence properties observed elsewhere for NN-only, 
with a net induced four-body contribution in $A=4$ that is smaller
than the truncation errors of the chiral interaction.    
The soft SRG interactions are
an alternative to the use of Lee-Suzuki effective interactions in
NCSM and the HO matrix elements can also be used (after conversion to a
Slater-determinant HO basis as needed) for coupled cluster and many-body
perturbation theory calculations. 
A more complete analysis of convergence and
dependencies for the energy and other observables for few-body
systems, as well as results for other interactions and choices of
generator in Eq.~\eqref{eq:flow}, will be given in a forthcoming
publication~\cite{Jurgenson:2009}.

\begin{acknowledgments}
We thank E.~Anderson, S.~Bogner, J.~Drut, R.~Perry, S.~Quaglioni, and 
A.~Schwenk for useful comments.  This work was
supported in part by the National Science Foundation under Grant
No.~PHY--0653312 and the UNEDF SciDAC Collaboration under DOE Grant
DE-FC02-07ER41457.
Prepared in part by LLNL under Contract DE-AC52-07NA27344.

\end{acknowledgments}


\end{document}